# IMAGE COMPRESSION WITH ENCODER-DECODER MATCHED SEMANTIC SEGMENTATION

Trinh Man Hoang[1], Jinjia Zhou[1,2], YiboFan[3]

[1]Graduate School of Science and Engineering, Hosei University, Tokyo, Japan
[2]JST, PRESTO, Tokyo, Japan
[3]Fudan University, Shanghai, China

## ABSTRACT

In recent years, deep learning has achieved great success in image compression. Especially, the layered compression is demonstrated to be a promising direction, which encodes a compact representation of the input image and apply an up-sampling network to reconstruct the image. To further improve the quality of the reconstructed image, some works transmit the semantic segment together with the compressed image data. But the compression ratio is also decreased because extra bits are required for transmitting the semantic segment. To solve this problem, we propose a new layered image compression framework with encoder-decoder matched semantic segmentation (EDMS). A semantic segmentation network is applied to the up-sampled image in both encoder and decoder. And then, followed by the semantic segmentation, a special convolution neural network is used to enhance the inaccurate semantic segment. As a result, the accurate semantic segment can be obtained in the decoder without requiring extra bits. The experimental results show that the proposed EDMS framework can get up to 2.5dB better PSNR than the HEVC-based (BPG) codec, 5% bitrate and 24% encoding time saving compare to the state-of-the-art semantic-based image codec.

## 1. INTRODUCTION

The typical lossy image encoding standards such as JPEG, JPEG2000 or HEVC-based BPG codec[1] are mostly processed based on block-wise transformation and quantization. In the case of limited transmission bandwidth, the large quantization parameter is usually assigned to achieve low bit-rate coding. However, it also leads to extreme blurring and block-type artifacts.

Over the past few years, deep learning has achieved great success in computer vision and image processing tasks. Especially, *Cheng et al* [2] have demonstrated that several learning-based methods like autoencoder, super-resolution and GAN[3] can be applied to perform the image compression task. Different from traditional hand-crafted block-based coding, deep learning-based approaches have the chance to extract and utilize the features of the data; therefore, they could get a better compression while avoiding block-type artifacts.

Many works of the deep learning-based image compression, which could outperform the traditional approach, were replied on the use of additional information. This additional information is varied from semantic information [4] [5], attention map[6] to a low-dimension version[7] of the image. Recently, thanks to the rapid evolving of the semantic segmentation technique, several techniques have achieved high performance whose results can be used in other tasks [8]. With the rapid development of GAN[3] in image synthesis using the semantic segment[9], the semantic segment has become the most common additional information in the field.

Several learning-based image compression methods have used semantic information to enhance their coding efficiency. *Liu et al*[5] designed a semantic analysis branch to guide their auto-encoder image codec but there is no experiment to demonstrate the impact of this branch, and the output from their analysis branch was not directly used to enhance the decoded image.

The state-of-the-art semantic-based image compression framework – DSSILC[4] sent the down-sampled image and the semantic segment to the decoder. DSSILC then up-sampled the compact image and used the GAN-based image synthesis technique[9] to reconstruct the image using the sending semantic segment and this up-sampled version. With the help of the semantic segment, DSSLIC could reconstruct better image quality than all the existed traditional image compression methods. However, DSSLIC requires the encoder to send extra bits of the semantic segment extracted from the original image to the decoder under a lossless compression.

To address this issue, we propose a new layered image compression framework with encoder-decoder matched semantic segmentation (EDMS). A semantic segmentation network is applied to the up-sampled image in both encoder and decoder. Hence, the same semantic segment can be obtained in decoder without extra bits. But the quality of the reconstructed image is decreased because the semantic segment extracted from the up-sampled image is not as accurate as of that from the original image. To obtain this quality gap, a convolution neural network (CNN) with a special structure is further applied to non-linear map the extracted segment to its original distribution before performing the image synthesis process. Experimental results show that our approach can get better performance than the state-of-the-art segmentation-based image compression. Furthermore, it does not only fill the mentioned gap but even reduces the encoding time.

## 2. PROPOSED IMAGE COMPRESSION FRAMEWORK: EDMS

We propose a new layered image compression framework with encoder-decoder matched semantic segmentation (EDMS). Based on EDMS, the decoder can use the semantic segment to enhance the quality of the reconstructed image without any extra bits.

### 2.1 Analysis of the segmentation in image codec

We first consider the way to get the semantic segment without sending it in the bitstream and design an extra branch to perform the same semantic extraction manner in both encoder and decoder. To fill out the distortion of the semantic segmentation, we furthermore use the Recursive Residual architecture to enhance the inaccurate semantic segment in our design.

The layered image compression systems usually encode and send the compact version of the original images to the decoder. In the decoder side, a super-resolution neural network is applied to reconstruct the compacted image. In order to improve the quality of the reconstructed image, the semantic segment is also sent as a piece of evidence for the reconstruction task. Because of that, a noticeable number of bits are used to store this evidence.

There will be several problems if we discard the semantic segmentation information and then perform it only at the decoder-end:

- If only performing segmentation on the up-sampling version of the image in the decoder side, the quality of the reconstructed image will be not good since the received residual was conducted based on the original semantic segment. We solve this problem with our synthesis idea in Section 2.2.
- Since the decoded image contains noise artifacts from the lossy compression, performing the segmentation directly on the decoded image will lead to inaccurate boundary decisions because of those artifacts. Therefore, we proposed a semantic enhancing network to keep the quality unchanged, the detail is shown in Section 2.3.

### 2.2 Encoder-decoder matched semantic segmentation

*Akbari et al*[4] introduced a CompNet at the encoder to down-sample the original image to a compact version. Especially,

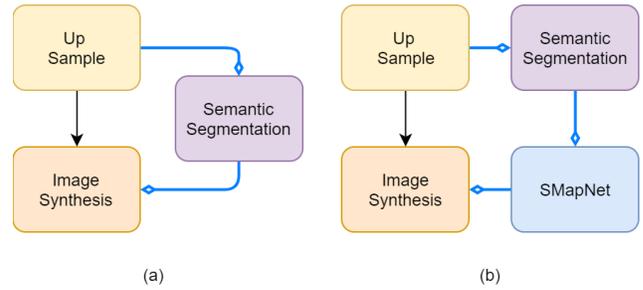

**Fig. 1.** (a) Perform segmentation on extra branch, synchronized for both encoder and decoder; (b) Apply SMapNet for segmentation enhancement before the image reconstruction

this network used the semantic segment as side information to perform the down-sampling. Since this compact version is conducted based on the semantic segment and usually lossless sent to the decoder, it is suitable for our seeking to perform the semantic segmentation in both sides. Besides, a GAN-based FineNet[9] was also demonstrated that it can generate a synthesis image from the up-sampled version and the semantic segment.

However, to leverage that useful compact version and powerful generative ability of GAN, we first trained the CompNet and FineNet as the down-sampling network and the image synthesis network, respectively, several times with the original semantic segment. It is not only to let the CompNet strong enough but it also for the growth of the image synthesis network-FineNet to leverage the up-sampled version and semantic segment.

After several training iterations, we extract the semantic segment from the up-sampled version of the compact image, then replace it as the input for the image synthesis network instead of the original semantic segment (see Fig 1a). The new synthesis image then is utilized to calculate the residual between it and the original image. Since this process can be repeated at the decoder, there is a correlation between the residual and the synthesis image without sending the semantic segment.

### 2.3 Semantic segmentation enhancement

The semantic segment from an extra branch is not as good as the one extracted from the original image. Since the down-sampled version is lack of details, the segmentation algorithm has less information to make the decision, especially for the deformation case. Therefore, we need a mapping process to map the deformation semantic to its true distribution.

By wondering about the special type of mapping information, the semantic segment is much simpler than a general image. It means that it will be easy to fall into the overfitting situation or gradient exploding when training. Hence, we applied the Recursive Residual architecture [10] as a mapping operator for this type of semantic information.

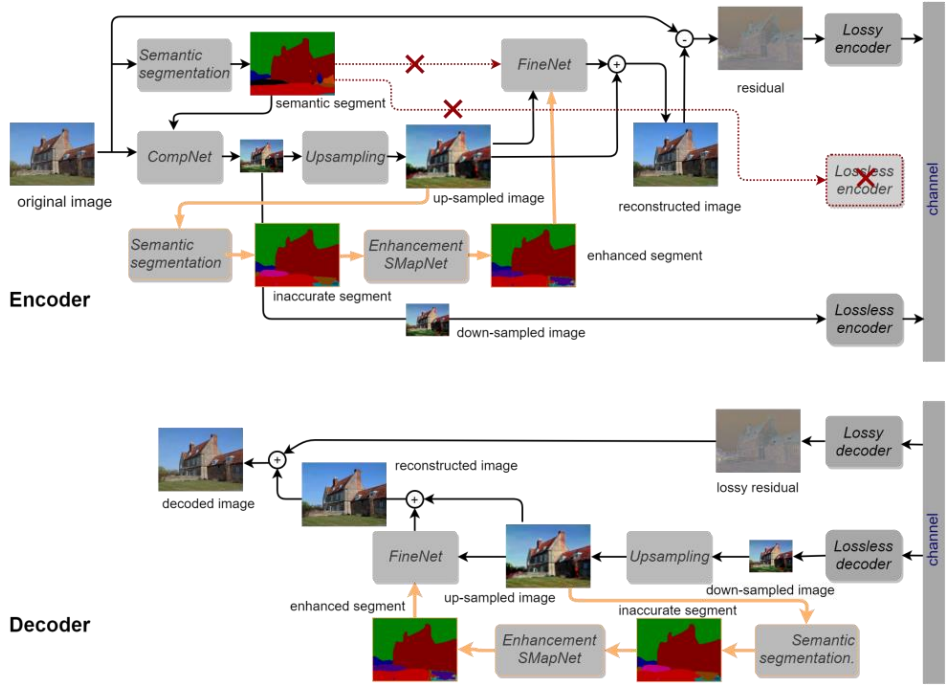

**Fig. 2.** Our proposed framework - EDMS with extra branch for segmentation enhancement

This architecture is demonstrated as a strong design again the overfitting issue and more stable for training than normal recurrent CNN.

The enhancing model is then applied for the inaccurate semantic segment from the up-sampled image - SMapNet (see Fig 1b). The residual now is calculated based on the synthesis image which conducted from the enhanced segment and the up-sampled image.

### 2.4 Overall framework

Fig 2 shows our overall framework, note that the original semantic segment is not sent to the decoder and only used in the training process.

On the encoder side, we first need to perform the down-sampling and up-sampling processes with the original semantic segment to obtain the semantic-driven up-sampled version. Next step, we extract the segment from this version, and use the SMapNet for semantic segmentation enhancement and input the SMapNet's output into the position of the semantic segment in the FineNet. The final residual will be calculated based on the output of FineNet forward with SNetMap segment as its input (see Fig 2). This residual then will be encoded by BPG[1] (state-of the-art traditional lossy image codec), lossless FLIF codec[11] is applied for the compact version of the image and there is no extra bit used for transferring the semantic segment.

On the decoded side, we received only the down-sampled image and lossy residual from the channel. The semantic segment used to reconstruct the decoded image is conducted from the up-sampled image and enhanced by our SMapNet. Next, FineNet uses this enhanced segment and the up-sampled image as it input to perform the reconstruction. Since we also performed this process on the encoder side, there always is a correlation between the received residual and this reconstructed image. The reconstructed image is then sum up with the residual to output the final decoded image.

There are three main networks in our framework: CompNet, FineNet (leverage from [4]) and SMapNet (proposed in this work). The architecture of these networks are as follows:

- CompNet: $^7c_{64}$, $^3c_{128}$, $^3c_{256}$, $^3c_{512}$, $^7c_3$, tanh; FineNet: $^7c_{64}$, $^3c_{128}$, $^3c_{256}$, $^3c_{512}$, 9 x $^3r_{512}$, $^3u_{256}$, $^3u_{128}$, $^3u_{64}$, $^7c_3$, tanh
- SMapNet: $^3c_{64}$, 9 x $^3v_{64}$, $^3c_3$ – no ReLu

where

- $^sc_k$: (s x s) convolution layer with k filters and stride 1, followed by instance normalization and ReLU.
- $^sr_k$: residual block with two (s x s) convolution layers and k filters, followed by instance normalization
- $^su_k$: (s x s) fractional-strided-convolution layer with k filters and stride ½, followed by instance normalization and ReLU.
- $^sv_k$: recursive residual block with two (s x s) convolution layers and k filters, followed by ReLU.

**Training procedure.** As we mentioned in Section 2.2, to obtain a good enough semantic segment from the up-sampled image, we first need to perform the training process based on the original semantic segment. The ADE20K [12] with 150 semantic labels are used for this stage of training. All the images are rescaled to 256x256 to have a fixed size of training. We set the down-sample factor of the CompNet equal to 8 to get the compact representation of size 32x32x3. Then we perform the training process with five kinds of losses according to [4], which is demonstrated that got better results on both perceptual and PSNR quality (see Step 1 in Fig 3).

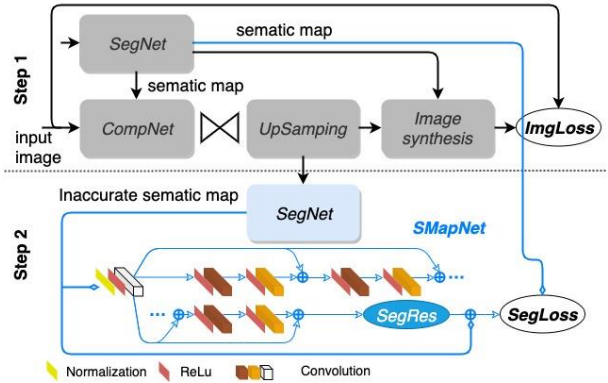

**Fig.3.** The specific training procedure

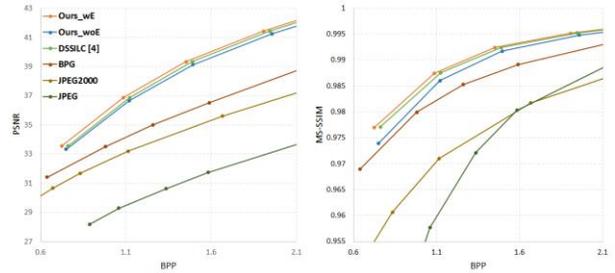

**Fig.4.** The comparison of compression techniques using PSNR (left) and MS-SSIM (right) on the ADE20K test set.

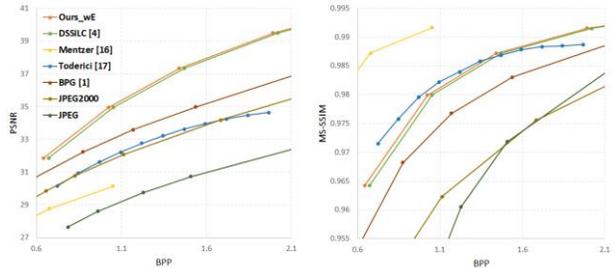

**Fig.5.** The comparison of compression techniques using PSNR(left) and MS-SSIM(right) on the Kodak test set with different training dataset

On the next step, based on the above pre-trained model, we use the PSP network[8] to perform the semantic segmentation on the up-sampling images (256x256 from the ADE20K dataset) and use them as inputs for training SMapNet. The SMapNet is trained as a non-linear mapping operator between the extracted segment and its original version (see Step 2 in Fig 3). For SMapNet, we use a minibatch size of 32 and Adam[13] optimizer, we start with a learning rate of 5e-04, for stable training, the final layer will have a learning rate equal to one tenth of other layers, then terminate training at 150 epoch.

## 3. EXPERIMENTAL RESULTS AND COMPARISON

### 3.1 Experiment settings

Our experiments were conducted on an NVIDIA Tesla V100 GPU while an Intel Core i7-8700K CPU was used to perform non-GPU tasks. We trained our models on the ADE20K dataset as we mentioned in the training procedure – Section 2.4. ADE20K[12] test set and Kodak[14] dataset are used as testing sets in our experiments. For the ADE20K dataset, we randomly select 50 test images that are not relevant to the training set. The results are recorded by average value over all test images. We use both PSNR and MS-SSIM [15] metrics to evaluate our results.

### 3.2 Evaluating the overall performance

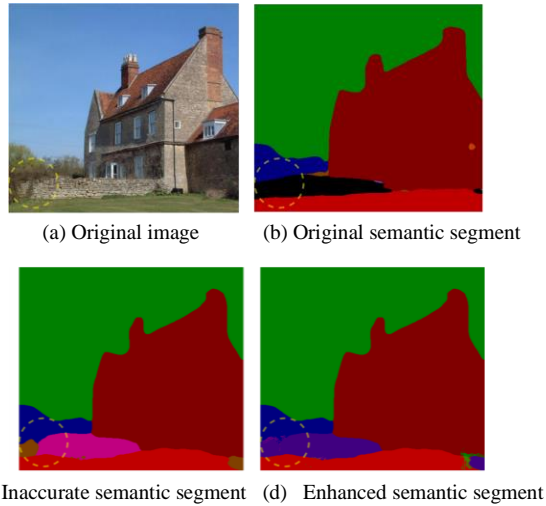

(a) Original image    (b) Original semantic segment

(c) Inaccurate semantic segment    (d) Enhanced semantic segment

**Fig.6.** Our SMapNet performance in the semantic enhancement task. Original sematic map is extracted from the original image. Inaccurate semantic segment is extracted from the up sampled image. Enhanced semantic segment is obtained by applying SMapNet to the inaccurate semantic segment.

**Performance gain.** As shown in Fig. 4, we compare the performance of our codec with JPEG, JPEG2000, H.265/HEVC intra-coding based BPG codec[1] and DSSLIC codec[4], which is state-of-the-art in semantic-based image compression. For a fair comparison, our residual is compressed by the BPG codec with the same QPs as the DSSLIC codec. Note that, in all figures and following discussion, Ours_woE and Ours_wE represent our approach with and without applying SMapNet respectively. As shown in Fig 4, our method – Ours_wE gets the upper bound over all other methods in both PSNR and MS-SSIM. Compare to the state-of-the-art semantic-based image codec DSSLIC, our method could gain 0.5-1dB when the bitrate (bits per pixel-bpp) is less than 1.1 and reduce 5% bitrate at the same PSNR. Comparing to the most efficient traditional image codec – BPG [1], our average PSNR is more than 2.5dB on the ADE20K test images with the bitrate in range of 1.1-2.1 bpp.

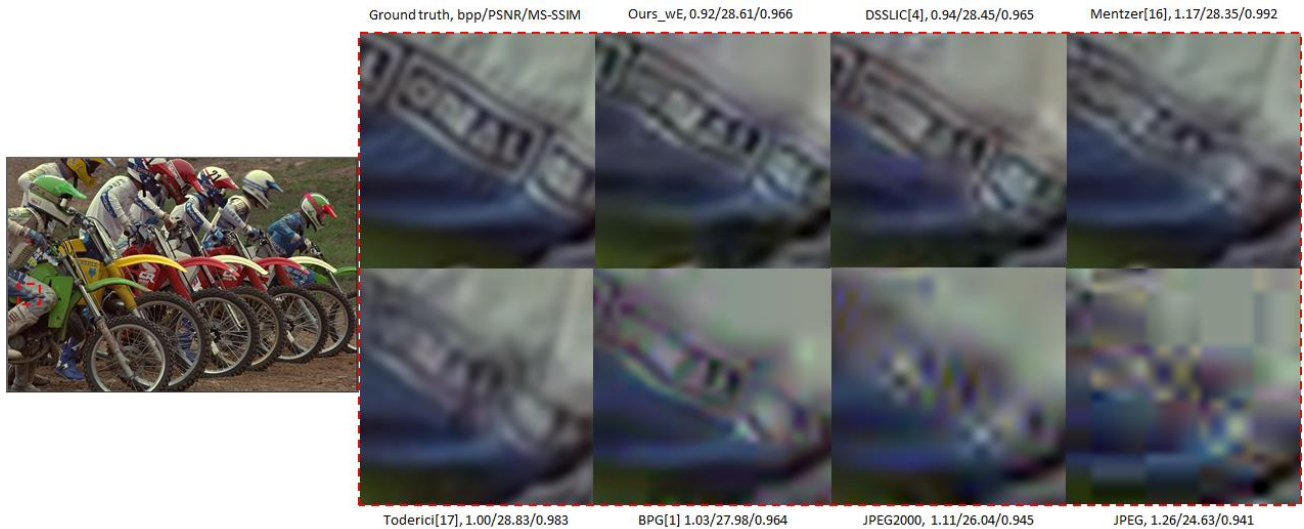

**Fig.7.** Qualitative comparison between different compression codec by bpp/PSNR/MS-SSIM. Note that our proposed method gets the best decoded quality with the smallest bitrate.

Table 1. Comparison of the compression quality and processing time (under the same BPG quantized parameter – QP = 32)

| Dataset | DSSLIC[4] | | | Ours: w/o Semantic Enhancement | | | Ours: w/ Semantic Enhancement | | |
|---|---|---|---|---|---|---|---|---|---|
| | bpp | PSNR (dB) / MS-SSIM | Enc./Dec. Time (s) | bpp | PSNR (dB) / MS-SSIM | Enc./Dec. Time (s) | bpp | PSNR (dB) / MS-SSIM | Enc./Dec. Time (s) |
| **ADE20K** | 0.762 | 33.57 / 0.977 | 0.838 / 0.315 | 0.75 | 33.33 / 0.974 | 0.579 / 0.247 | 0.726 | 33.57 / 0.977 | 0.709 / 0.377 |
| **Kodak** | 0.671 | 31.86 / 0.964 | 1.08 / 0.342 | 0.657 | 31.77 / 0.959 | 0.703 / 0.2635 | 0.642 | 31.87 / 0.964 | 0.745 / 0.305 |

**Semantic segment enhancement.** To demonstrate the effect of enhancing the semantic segmentation, we record the results without using it – Ours_woE (see Fig 4, 6 and Table 1). Since the up-sampled image is lack of details, the inaccurate semantic segment does not perform very well, however, it still gets better results when comparing to all other traditional methods. Fig 6 shows the enhancing effect of our SMapNet, we can see clearly a wrong split wall in the raw segment has been connected again by SMapNet.

**Processing speed.** Furthermore, we also compare the encoding and decoding time among DSSLIC, Ours without Enhancement and Ours with Enhancement models in Table 1.

We could clearly see that without semantic segment compression, Ours_woE method could perform much faster than DSSLIC but it reduces the quality of decoded image in trade-off. With the participation of SMapNet, Ours_wE method could fill the quality gap with smaller bpp, reduce 24% encoding time and almost unchanged the decoding time.

**Generalization capability.** In order to demonstrate the generalization capability of our method, we further test our methods on the Kodak dataset[14]. Our ADE20K-trained model is used to perform the test on 24 Kodak images. We compare our method with some learning-based image codecs like DSSLIC, Conditional Probability models from Mentzer [16], Toderici's work[17] and several traditional methods like BPG[1], JPEG2000 and JPEG. The average results are visualized in Fig 5. From Fig 5, we can observe that our method still achieves the upper-bound in PSNR on the Kodak test set. In particular, the average PSNR gained in range 1.3-2.3 bpp is 2.4dB better for our methods when comparing to the BPG codec. This result demonstrates that our method generalizes well when the training and testing images are from different distributions. Note that, for Mentzer's and Toderici's works, since the provided models were designed and trained by MS-SSIM loss, it is easy to understand their poor performance on PSNR and the opposite results on MS-SSIM.

**Subjective evaluation**. A visual example from the Kodak dataset is shown in Fig 7. Our method could get the best image quality with the smallest bpp. When looking into the cropped part, we could clearly see that JPEG and JPEG2000 got a lot of block artifacts and noise. While BPG, Mentzer's and Toderici's models smoothed over some parts. And DSSILC needed more bits to reconstruct an image with similar quality as ours.

## 4. CONCLUSION

This paper presents a novel layered image compression framework for leveraging the semantic segment without transferring any extra bit. It is challenging to keep the quality of decoded images while saving the bits for transferring the semantic segment. These challenges come from the position to perform the segmentation, the synthesis of that process between encoder and decoder, how to reduce the distortion of extracted segmentation, and so on. With our idea of encoder-decoder matched semantic segmentation (EDMS), semantic segment enhancement and specific training procedure, our model could get over all the challenges.

Experimental results showed that the proposed approach could outperform all traditional image codecs, including H265/HEVC-based BPG, JPEG2000 or JPEG in terms of PSNR and MS-SSIM metric. Furthermore, at the same PSNR, it could gain up to 5% bitrate and 24% encoding time reduction compare to DSSILC[4], the state-of-the-art semantic-based image codec. Since there still have a lot of information can be synchronously extracted from both encoder and decoder, our approach is potential to be applied to other future work.